\documentclass[prd,twocolumn,showpacs,a4paper]{revtex4}

\begin{document}

\title{Holography and the Cosmic Coincidence}

\author{Saulo Carneiro}

\affiliation{Instituto de F\'{\i}sica, Universidade Federal da
Bahia, 40210-340, Salvador, BA, Brazil}

\begin{abstract}
Based on an analysis of the entropy associated to the vacuum
quantum fluctuations, we show that the holographic principle,
applied to the cosmic scale, constitutes a possible explanation
for the observed value of the cosmological constant, theoretically
justifying a relation proposed $35$ years ago by Zel'dovich.
Furthermore, extending to the total energy density the conjecture
by Chen and Wu, concerning the dependence of the cosmological
constant on the scale factor, we show that the holographic
principle may also lie at the root of the coincidence between the
matter density in the universe and the vacuum energy density.
\end{abstract}


\maketitle

In spite of the great success of the standard model in describing
the evolution and structure of the universe, there are at least
two unsolved problems, both related to the existence of a
cosmological constant $\Lambda$. On the one hand, if $\Lambda$
originates from vacuum quantum fluctuations, its theoretically
expected value has the order of $l_p^{-2}$, where $l_p \equiv
\sqrt{\hbar G/c^3} \approx 10^{-35}$m is the Planck length
\cite{9}. That is, $122$ orders of magnitude greater than the
observed value $\Lambda \approx 10^{-52}$m$^{-2}$ \cite{8}. This
huge discrepancy is known as the cosmological constant problem.

On the other hand, in an expanding universe with scale factor
$a(t)$ ($t$ is the cosmological time), $\Lambda$ is a constant,
while the matter density $\rho_m$ falls with $a^3$. Therefore, it
is an important problem to understand why the observed matter
density is so close to the vacuum energy density. This second
question is usually referred to as the cosmic coincidence problem.

Another curious coincidence is the so-called large numbers
coincidence \cite{22,GS}, pointed out by Eddington in the
beginning of the 20th century, and which may be expressed in the
form
\begin{equation}
\label{EW} \hbar^2 H \approx G c m^3,
\end{equation}
where $m$ is a typical hadronic mass, and $H \equiv \dot{a}/a$ is
the Hubble parameter.

This surprising relation between physical quantities
characteristic of the micro and macrocosmos led Dirac to postulate
a possible variation of $G$ with the cosmological time, in such a
way that (\ref{EW}) would be valid during the entire evolution of
the universe \cite{4}. An alternative explanation, in the spirit
of the anthropic principle, was suggested by Dicke, who argued
that (\ref{EW}) is not valid for all times, being only
characteristic of our era \cite{Dicke}.

As we shall see, the large numbers coincidence and the problems
related to the cosmological constant are not, in fact, independent
problems. They can be related with the help of the holographic
principle, postulated some years ago by 't Hooft \cite{11} and
Susskind \cite{11'}. We will try to show that the application of
this principle to the universe as a whole leads to a value for the
cosmological constant in accordance with recent observations.
Furthermore, with the help of ideas originally proposed by Chen
and Wu \cite{Wu}, with respect to the dependence of the vacuum
energy density on the scale factor, we will show that the
holographic principle may also lie at the root of the cosmic
coincidence problem.

In a simplified form, this principle can be described as the
extension, to any gravitating system, of the Bekenstein-Hawking
formula for black-hole entropy. More precisely, it establishes
that the maximum entropy of a gravitating system has the order of
$A/l_p^2$, where $A$ is the area of a characteristic surface which
bounds the system, defined in an appropriate manner. In this way,
the number of degrees of freedom of the system is not bounded by
its volume in Planck units, as expected from quantum theories of
space-time, but by the area, in Planck units, of its delimiting
surface.

The use of this principle to the universe as a whole depends on
the definition of such a surface at a cosmic scale. The existence
of a cosmological constant naturally introduces a characteristic
surface of radius $\Lambda^{-1/2}$, which may be considered for
application of the holographic principle \cite{18}. Let us call
this version of the cosmological holographic principle (CHP) as
the weak version, for reasons that will soon be clear.

Another version of the CHP \cite{17}, referred to hereafter as the
strong version, uses as the characteristic bounding surface the
Hubble horizon, with radius $c/H$, for it defines the scale of
causal connections for any observer, or what we call the
observable universe. It is clear that in an homogeneous and
isotropic, infinite (open or flat) universe, filled with a perfect
fluid and a positive cosmological constant, the strong version
implies the weak one, for the Hubble radius tends asymptotically
to $\sqrt{3/\Lambda}$.

In a previous work \cite{GS}, the CHP is used, in its weak
version, as the basis of a possible explanation for the origin of
relation (\ref{EW}). To do that, the number of degrees of freedom
in the universe is identified with the number of barionic degrees
of freedom. Even without entering into the details of the
contribution of non-barionic dark matter to the entropy of the
universe, we know that this assumption does not have observational
basis, since the cosmic background radiation contains about
$10^{8}$ photons per barion. Moreover, the major contribution to
the entropy of matter seems to come from massive black-holes
present in galactic nuclei, which represent an entropy (for
$\kappa_B=1$) of the order of $10^{101}$, while the number of
barions in the universe is of the order of $10^{80}$
\cite{Penrose}.

This difficulty can be overcome if we observe that the entropy of
the universe is dominated by the vacuum quantum fluctuations that
lead, in the classical limit, to the cosmological constant.
Following Matthews \cite{RM}, if we suppose that the vacuum energy
density in each cosmological epoch originates from a phase
transition with characteristic energy scale $m c^2$, we can
associate to this scale a Compton length $l \approx \hbar/(mc)$.
Therefore, the correspondent number of observable degrees of
freedom will be of the order of $V/l^3$, where $V$ is the volume
of the observable universe. In other words,
\begin{equation}
\label{N} N \approx \left(\frac{R}{l}\right)^3 \approx
\left(\frac{m c^2}{\hbar H}\right)^3
\end{equation}
($R$ is the Hubble radius).

The last phase transition resulted from quark-hadron confinement,
with characteristic energy $m c^2 \approx 150$ MeV \cite{RM}.
Taking for $H$ the value observed nowadays, $H \approx 65$
km/(sMpc) \cite{8}, we obtain $N \approx 10^{122}$, a value that
indeed predominates over the entropy of matter given above, of
order $10^{101}$. Actually, it is easy to verify that (\ref{N})
dominates the entropy of the universe for any $R$ above the value
of decoupling between matter and radiation.

On the other hand, the CHP establishes that the maximum number of
degrees of freedom available in the universe is given by
\begin{equation}
\label{Nmax} N_{max} \approx \left(\frac{R}{l_p}\right)^2 =
\left(\frac{c}{H l_p}\right)^2.
\end{equation}

If we are restricted to the weak version, we should take for $H$
the asymptotic value $c\sqrt{\Lambda/3}$. In the same way, the
maximum value of (\ref{N}) corresponds to this asymptotic value as
well. Identifying both results, we then have
\begin{equation}
\label{Lambda} \Lambda \approx G^2 m^6/\hbar^4.
\end{equation}

This expression was derived $35$ years ago by Zel'dovich
\cite{Zel'dovich}, from empirical arguments. It was recently
re-derived by Matthews \cite{RM}, who takes relation (\ref{EW}) as
given, as well as the present dominance of the cosmological
constant over the density of matter.

As observed by Matthews, (\ref{Lambda}) is a remarkable relation.
Although it is sensible to the parameter $m$, it leads to correct
orders of magnitude, whatever the transition we consider: that of
the inflationary epoch, the electroweak transition, or the latest
one, due to quark-hadron confinement. In this last case, using
$mc^2 \approx 150$ MeV, we obtain $\Lambda \approx 10^{-51}$
m$^{-2}$, in good agreement with observation, considering that we
are not taking into account numerical factors.

Now, if we assume that the present evolution of the universe is
dominated by the cosmological constant, as corroborated by
observation \cite{8}, we can set $H \approx c\sqrt{\Lambda}$, and
(\ref{Lambda}) reduces to (\ref{EW}), the origin of which is then
explained. Note that, here, this relation is valid only
asymptotically, that is, for times when the cosmological constant
dominates. In this respect, the explanation derived via the weak
CHP has the same character as the anthropic solution given by
Dicke.

On the other hand, if we consider Dicke's solution valid,
therefore justifying (\ref{EW}), the result (\ref{Lambda}) implies
the present dominance of the cosmological constant, which can be,
in this way, understood \cite{GS}. We see therefore that the weak
CHP incorporates a new theoretical ingredient to previous
approaches, for it represents an independent explanation for
relation (\ref{Lambda}).

Let us consider now the strong version of the CHP. As we saw,
after the decoupling between matter and radiation, the entropy
(\ref{N}) associated to the vacuum quantum fluctuations dominates
the total entropy of the universe. On the other hand, it is easy
to see that (\ref{N}) represents also the maximum value (for a
given $R$) of the vacuum entropy. Therefore, the strong version of
the CHP implies the identity between (\ref{N}) and (\ref{Nmax}),
leading to $\hbar^2 H \approx Gcm^3$. That is,  it leads directly
to the large numbers coincidence, eq. (\ref{EW}), without any
additional assumption as the dominance of $\Lambda$ or anthropic
arguments.

The price to pay is that, now, (\ref{EW}) is valid for any
cosmological time, as originally proposed by Dirac. As a
consequence, $G$ will vary with $R$ according to
\begin{equation}
\label{G} G \approx \frac{(\hbar^2/m^3)}{R} = \left(
\frac{\hbar^2}{cm^3} \right) H.
\end{equation}

This corresponds to the relative variation
\begin{equation} \label{G/G}
\dot{G}/G = \dot{H}/H = - (1+q) H,
\end{equation}
where $q \equiv -a\ddot{a}/\dot{a}^2$ is the deceleration factor.
Later on we shall discuss the compatibility of this result with
the bounds imposed by observation on a possible time variation of
$G$. For now, let us observe that, in a universe dominated by the
cosmological constant, $H \approx c\sqrt{\Lambda}$ is a constant,
and $q \approx -1$. In this case, (\ref{G}) and (\ref{G/G}) lead
to (\ref{Lambda}), with $G$ constant, as expected on the basis of
the weak version of the CHP.

Let us set aside for now the holographic principle and its
cosmological consequences, to introduce some ideas originally
proposed by Chen and Wu \cite{Wu}, concerning the dependence of
the vacuum energy density on the scale factor. The variation of
$\Lambda$ with $a$ was suggested by those authors as a solution to
the problem of the cosmological constant, the value of which would
relax with the expansion of the universe. If we associate
$\Lambda$ with the vacuum quantum fluctuations, in the Planck time
its value had the order of $l_p^{-2}$. Therefore, if we assume
that $\Lambda$ varies with $a$ as a power law, such a dependence
should have, modulo a constant factor of the order of unity, the
form
\begin{equation}
\label{Wu} \Lambda \approx l_p^{-2} \left(\frac{l_p}{a}\right)^n.
\end{equation}

The essential argument pointed out by Chen and Wu is that, if we
expect a classical evolution for large times, there should not be
any dependence on the Planck constant in the above relation. In
this way we should have $n=2$, and therefore $\Lambda \approx
a^{-2}$. For $a=l_p$, we obtain evidently $\Lambda \approx
l_p^{-2}$. But, for the present value  $a \approx 10^{-26}$m, one
has $\Lambda \approx 10^{-52}$m$^{-2}$, in accordance with
observation.

The weakness of the above reasoning is that, instead of taking
exclusively an explicit dependence on $a$, as we have done, we can
consider also a dependence on, for example, the Hubble radius
$R=\dot{a}/a$, or still on the cosmological horizon. The different
horizons do not have, in general, the same order of magnitude, and
their relation with $a$ depends on the metric, that is, on the
complete matter content considered. As we shall see below, this
difficulty is naturally overcome when we extend the conjecture of
Chen and Wu to the total energy density, as proposed by M.V. John
and K.B. Joseph \cite{JJ}.

Let us suppose that the universe starts with a quantum fluctuation
at the Planck scale. The matter content generated in this way will
have an initial energy density given by $G \rho/c^4 \approx
l_p^{-2}$, which, by the same arguments exposed above, will fall
with $a$ by the same law (\ref{Wu}), with $n=2$. That is, the
total energy density will obey the conservation law
\begin{equation}
\label{JJ} \rho a^2 \approx c^4/G.
\end{equation}

It is easy to verify that, in an isotropic and homogeneous
space-time, such a conservation law leads to a linear expansion,
with scale factor $a \approx ct$ and $q=0$. We then have $H
\approx c/a$, and so $\rho \approx H^2c^2/G$. Thus, one sees that
the total energy density has the order of the critical density,
$\rho_c = 3H^2c^2/(8\pi G)$, in accordance with the observed
flatness of universe \cite{8}.

We see as well that the Hubble radius $R = c/H$ coincides with the
scale factor. Therefore, the question of which of the two should
be considered for expressing the variation of $\rho$ through
equation (\ref{Wu}) is irrelevant here. The same can be said with
respect to the cosmological horizon, for in an approximately flat
universe it has the same order of magnitude as the Hubble horizon.

This is a reasonable result, which shows the internal consistency
of this approach. In an isotropic and homogeneous universe born
from a quantum fluctuation, the only available parameter, besides
the fundamental constants $G$, $c$ and $\hbar$, is the scale
factor, which has a purely geometric nature. Therefore, the matter
content so created should lead, necessarily, to causal horizons of
the same order of $a$.

Now, if we treat our matter content as a perfect fluid, the
conservation law (\ref{JJ}), with $G$ constant, corresponds to the
equation of state $p=-\rho/3$. Furthermore, considering it a
bi-component fluid \cite{JJ}, composed of dust and of a
cosmological constant with energy densities $\rho_m$ and
$\rho_{\Lambda}$, respectively, we have
\begin{eqnarray}
\label{eqn} \rho = \rho_m + \rho_{\Lambda}, \nonumber \\ p =
p_{\Lambda} = - \rho_{\Lambda}.
\end{eqnarray}

Note that, in equations (\ref{eqn}), $\rho_m$ and $\rho_{\Lambda}$
do not evolve independently, because of the constraint imposed by
the conservation law (\ref{JJ}). This constraint leads to a small
production of matter (coming from the decaying vacuum energy
density), which, as shown in reference \cite{JJ}, is far beyond
the current possibilities of detection.

Substituting in relations (\ref{eqn}) the equation of state given
above, we obtain $\rho_m = 2 \rho_{\Lambda}$. Although a linear
expansion, with a null deceleration factor, is compatible with the
latest observations \cite{JJ}, such a relation between $\rho_m$
and $\rho_{\Lambda}$ is not in accordance with the observational
data \cite{8}. As we shall see below, this crucial problem in the
approach by John and Joseph can be solved with the help of the
holographic principle, leading to results in remarkable agreement
with observation.

In a context in which $G$ varies with $a$, as predicted by the
strong version of the CHP, it is possible to show that the
conservation law (\ref{JJ}) leads as well to a uniform expansion,
with $a \approx ct$. However, using (\ref{G}) (with $R \approx
a$), equation (\ref{JJ}) turns out to be
\begin{equation}
\label{rho} \rho a \approx c^4 m^3/\hbar^2.
\end{equation}

That is, the total energy density does not fall with $a^2$, as
before, but with $a$. Therefore, the continuity equation leads now
to a new equation of state, given by $p = -2\rho/3$. Substituting
in (\ref{eqn}), we obtain $\rho_{\Lambda} = 2\rho_m$. In
particular, for $\rho=\rho_c$ this gives $\rho_m/\rho_c = 0.33$,
and $\rho_{\Lambda}/\rho_c = 0.67$.

Note that these are exact results, for they do not depend on
numerical factors that we have been neglecting throughout this
paper. We thus have an excellent accordance with observation
\cite{8}. It constitutes a natural explanation for the cosmic
coincidence problem, based on simple and fundamental ingredients:
the cosmological holographic principle, in its strong version, and
considerations about the conservation of energy in an isotropic
and homogeneous universe born from a quantum fluctuation at the
Planck scale.

Evidently, expression (\ref{Lambda}), which gives the absolute
value of $\Lambda$ for a given value of $G$ (and, in particular,
the value observed nowadays), remains valid. Indeed, we have seen
that $\rho_{\Lambda} \approx \rho \approx H^2c^2/G$. On the other
hand, by definition, $\rho_{\Lambda} \approx \Lambda c^4/G$.
Equating both results, we have $\Lambda c^2 \approx H^2$, which,
substituted in (\ref{G}), leads directly to equation
(\ref{Lambda}).

We should not forget that these results refer to late eras of
universe evolution, i.e., after the decoupling between matter and
radiation, when the expansion is dominated by a bi-component
fluid, composed of dust and of a cosmological constant. For early
times, all will depend on the matter content considered. In
addition, in early eras the major contribution to the total
entropy does not come from vacuum quantum fluctuations, which
changes the reasoning developed here.

This problem does not belong to the scope of this article, but let
us consider, just as an example, a universe dominated by radiation
at temperature $T$. If we associate to each photon an energy
$mc^2$, and the corresponding Compton length $l \approx
\hbar/(mc)$, the maximum number of degrees of freedom remains
given, for each time, by (\ref{N}). We can then re-obtain
(\ref{rho}) by means of the same arguments used before.

But now $m$ is not constant anymore, varying with the temperature
as $mc^2 \approx \kappa_B T$, where $\kappa_B$ is the Boltzmann
constant. On the other hand, we know that, for radiation in
thermal equilibrium at temperature $T$, $\rho \propto T^4$.
Incorporating both results in equation (\ref{rho}), we obtain
$aT=\;$constant, or yet $\rho a^4=\;$constant. Thus, we see that
the CHP, in its strong version, together with the conjecture by
John and Joseph, is consistent with the conservation law for
radiation characteristic of the standard model.

This analysis allows us to understand the present dominance of the
entropy associated to the vacuum quantum fluctuations over the
radiation entropy, which we referred to at the beginning. The
former is given by (\ref{N}), with $m$ constant, i.e.,
$N_{\Lambda} \propto (a/\dot{a})^3$. On the other hand, the
maximum entropy of radiation, $N_{\gamma}$, is also given at each
instant by (\ref{N}), but with $m$ increasing with $T$, that is,
falling with $a$. We then have $N_\gamma \propto \dot{a}^{-3}$,
which leads to $N_{\Lambda}/N_{\gamma} \propto a^3$. Thus, we see
that, for small values of $a$, $N_{\gamma}$ dominates, while the
contrary occurring for later eras.

This result does not depend on the kind of expansion we are
considering (actually, it does not depend on the CHP, nor in the
weak nor in the strong versions). But, if the expansion is
uniform, as predicted by (\ref{JJ}), we see that $N_{\gamma}$ will
be constant over all the radiation era, as should be for a system
in thermodynamic equilibrium.

To conclude, let us finally discuss the compatibility of equations
(\ref{G}) and (\ref{G/G}) with the bounds imposed by observation
on a possible time variation of $G$. As we have seen, equation
(\ref{JJ}) leads to a uniform expansion, with $q=0$. From
(\ref{G/G}) we then have, at the present time, $\dot{G}/G = - H
\approx -7 \times 10^{-11}/$year.

Recent astronomical observations, generally based on measurements
of variations of relative distances in the solar system and in
binary pulsars, impose more restrictive limits on the relative
variation of $G$ \cite{5}. These results are, thus, in favor of
the weak version of the CHP. In this case, relation
(\ref{Lambda}), which explains the observed value of the
cosmological constant, would remain valid. The cosmic coincidence
problem, however, should wait for another solution.

Nevertheless, let us emphasize that such kind of observations
cannot give the ultimate answer to the question. We are
considering here a time variation of $G$ over a cosmic scale, in a
universe supposed spatially homogeneous and isotropic at the large
scale, but not locally. In this context, $G$ is a dynamical
quantity that can, at small scales, remain constant, and even
suffer spatial variations. Concluding, from local observations,
that $G$ does not vary over cosmological scales would be the same
as, for example, to conclude, based on observations restricted to
our galaxy, that the matter density in the universe is constant at
the large scale.

A definite conclusion with respect to this depends on establishing
precise limits by means of faithful cosmological observations. Up
to our knowledge, such kind of cosmological bounds still are not
sufficiently reliable to close the question. In any case, a
conclusive study on possible time variations of $G$ will permit
deciding between the two versions of the CHP we have been
considering.

${}$

I am greatly thankful to Prof. P. F. Gonzalez-Diaz, for his warm
hospitality and helpful suggestions during the two years I have
been visiting the IMAFF-CSIC, in Madrid. I am also grateful to G.
A. Mena Marug\'{a}n, L. Garay and M. Moles, from IMAFF, for useful
discussions, and to O. Pessoa Jr., from UFBA, for proofreading the
manuscript. This work was partially supported by CNPq, Brazil.

\end{document}